\documentclass[12pt,preprint]{aastex}

\usepackage{emulateapj5}
\submitted{Accepted for publication in ApJ Letters}

\begin{document}

\title{On The Spectrum and Spectropolarimetry of Type Ic Hypernova SN~2003dh/GRB~030329\footnotemark[1]}

\footnotetext[1]{Based on data obtained at the Subaru Telescope,
  which is operated by the National Astronomical Observatory of
  Japan (NAOJ)}

\author{
K.~S.~Kawabata\altaffilmark{2},
J.~Deng\altaffilmark{3,4},
L. Wang\altaffilmark{5},
P.~Mazzali\altaffilmark{3,4,6},
K.~Nomoto\altaffilmark{3,4},
K.~Maeda\altaffilmark{3,4},
N.~Tominaga\altaffilmark{3,4},
H.~Umeda\altaffilmark{3,4},
M.~Iye\altaffilmark{2,7},
G.~Kosugi\altaffilmark{8},
Y.~Ohyama\altaffilmark{8},
T.~Sasaki\altaffilmark{8},
P.~H\"oflich\altaffilmark{9},
J.~C.~Wheeler\altaffilmark{9},
D.~J.~Jeffery\altaffilmark{10},
K.~Aoki\altaffilmark{8},
N.~Kashikawa\altaffilmark{2},
T.~Takata\altaffilmark{8},
N.~Kawai\altaffilmark{11,12},
T.~Sakamoto\altaffilmark{11,12},
Y.~Urata\altaffilmark{11,12},
A.~Yoshida\altaffilmark{13},
T.~Tamagawa\altaffilmark{12},
K.~Torii\altaffilmark{12},
W.~Aoki\altaffilmark{2},
N.~Kobayashi\altaffilmark{14},
Y.~Komiyama\altaffilmark{8},
Y.~Mizumoto\altaffilmark{2,15},
J.~Noumaru\altaffilmark{8},
R.~Ogasawara\altaffilmark{8},
K.~Sekiguchi\altaffilmark{8},
Y.~Shirasaki\altaffilmark{15},
T.~Totani\altaffilmark{16}
J.~Watanabe\altaffilmark{17,2},
T.~Yamada\altaffilmark{2},
}

\altaffiltext{2}{Opt. \& IR Astron. Div., NAOJ, Mitaka,
 Tokyo 181-8588, Japan; koji.kawabata@nao.ac.jp,
 iye@optik.mtk.nao.ac.jp, kashik@zone.mtk.nao.ac.jp,
 aoki.wako@nao.ac.jp, mizumoto.y@nao.ac.jp,
 yamada@optik.mtk.nao.ac.jp.}
\altaffiltext{3}{Dept. of Astron., University of Tokyo,
Bunkyo-ku, Tokyo 113-0033, Japan; deng@astron.s.u-tokyo.ac.jp,
 mazzali@ts.astro.it, nomoto@astron.s.u-tokyo.ac.jp,
 maeda@astron.s.u-tokyo.ac.jp, ntominaga@astron.s.u-tokyo.ac.jp,
 umeda@astron.s.u-tokyo.ac.jp.}
\altaffiltext{4}{RESCEU, University of Tokyo,
  Bunkyo-ku, Tokyo 113-0033, Japan}
\altaffiltext{5}{Lawrence Berkeley National Laboratory, 1 Cyclotron
  Road, Berkeley, CA 94720, USA; lifan@panisse.lbl.gov.}
\altaffiltext{6}{INAF-Osservatorio Astronomico, Via Tiepolo, 11,
  34131 Trieste, Italy}
\altaffiltext{7}{Dept. of Astron., Graduate University
  for Advanced Studies, Mitaka, Tokyo 181-8588, Japan}
\altaffiltext{8}{Subaru Telescope, NAOJ, 650 North A'ohoku Place,
  Hilo, HI 96720, USA; george@naoj.org, ohyama@naoj.org, 
  sasaki@naoj.org, kaoki@naoj.org, takata@naoj.org,
  komiyama@naoj.org, noumaru@naoj.org, ryu@naoj.org, kaz@naoj.org}
\altaffiltext{9}{Dept. of Astron., University of Texas, Austin,
  TX 78712, USA; pah@alla.as.utexas.edu, wheel@astro.as.utexas.edu.}
\altaffiltext{10}{Dept. of Physics, New Mexico Institute of
  Mining and Technology, Socorro, NM 87801, USA; jeffery@kestrel.nmt.edu.}
\altaffiltext{11}{Dept. of Physics, Tokyo Institute of Technology, 
  Meguro-ku, Tokyo 152-855, Japan; nkawai@phys.titech.ac.jp, 
  sakamoto@hp.phys.titech.ac.jp, urata@crab.riken.go.jp.}
\altaffiltext{12}{RIKEN, Wako, Saitama 351-0198, Japan; 
  tamagawa@crab.riken.go.jp, torii@crab.riken.go.jp.}
\altaffiltext{13}{Dept. of Physics, Aoyama Gakuin University,
  Sagamihara, Kanagawa 229-8558, Japan; ayoshida@phys.aoyama.ac.jp.}
\altaffiltext{14}{Inst. of Astron., University of Tokyo, 
  Mitaka, Tokyo 181-0015; naoto@ioa.s.u-tokyo.ac.jp.}
\altaffiltext{15}{Astron. Data Analysis Center, NAOJ, Mitaka,
  Tokyo 181-8588, Japan; yuji.shirasaki@nao.ac.jp.}
\altaffiltext{16}{Dept. of Astron., Kyoto University, 
  Sakyo-ku, Kyoto 606-8502, Japan;
  totani@kusastro.kyoto-u.ac.jp.}
\altaffiltext{17}{Public Relations Center, NAOJ, Mitaka,
  Tokyo 181-8588, Japan; jun.watanabe@nao.ac.jp.}

\begin{abstract}

Spectroscopic and spectropolarimetric observations of
SN~2003dh/GRB 030329 obtained in 2003 May using the Subaru 8.2 m
telescope are presented. The properties of the SN are
investigated through a comparison with spectra of the Type Ic
hypernovae SNe 1997ef and 1998bw.
(Hypernovae being a tentatively defined class of SNe
with very broad absorption features: these features
suggest a large velocity of the ejected material and
possibly a large explosion kinetic energy.)
Comparison with spectra of other hypernovae shows that
the spectrum of SN~2003dh obtained on 2003
May 8 and 9, i.e., 34--35 rest-frame days after the GRB (for
$z=0.1685$), are similar to those of SN~1997ef obtained $\sim
34-42$ days after the fiducial time of explosion of that SN. 
The match with
SN~1998bw spectra is not as good (at rest 7300--8000 \AA ), 
but again spectra obtained $\sim
33-43$ days after GRB 980425 are preferred. This indicates that
the SN may have intermediate properties between SNe 1997ef and
1998bw.  Based on the analogy with the other hypernovae, the time
of explosion of SN~2003dh is then constrained to be between $-8$
and $+2$ days of the GRB.  The Si and O P-Cygni lines
of SN~2003dh seem comparable to those of SN~1997ef, 
which suggests that the ejected mass in SN~2003dh may match 
that in SN~1997ef.
Polarization was marginally detected at optical wavelengths. 
This is consistent with measurements of the late afterglow, 
implying that it mostly originated in the interstellar medium 
of the host galaxy.

\end{abstract}


\section{INTRODUCTION}

Following the discovery of the energetic and bright Type Ic
SN~1998bw in the error box of GRB~980425, the argument for the
association between at least some GRB's and some SNe Ic has become
a hot issue \citep{gal98,iwa98,kul98}. Other bright and energetic
SNe Ic have subsequently been discovered (e.g., SNe~1997ef
and 2002ap; see \citealt{nom01,bra01,maz02,nom03}), but in none of these
cases was there strong evidence for association with a GRB,
leaving doubts about the reality of the case of SN~1998bw and
GRB~980425.

The lack of association of all hypernovae with GRBs can
be explained by asymmetry.
GRBs could be produced by relativistic jets emerging from
asymmetric hypernovae (e.g., \citealt{wan98,iwa98,iwa00,
hoe99,woo99,nom01,nom03,whe01,mac01}).
It is generally assumed that the GRB is highly asymmetric
and collimated more or less closely along the jet axis.
There are also cumulative observational facts indicating that
Type Ib/Ic SNe are intrinsically asymmetric
(e.g., \citealt{wan01,maz01,mae02,mae03a,mae03b,wan03}).
Along with evidence for GRBs associated with some SNe,
there is evidence for SNe associated with some GRBs.
The bumps observed in some GRB afterglows may be
underlying SN light curves seen rising to
and passing maximum light \citep{blo02}.
However, spectral observations of the afterglows with bumps have not
been possible (until GRB 030329 and with the exception of 
GRB 021211; \citealt{mdv03}) because of their faintness.

The recent gamma-ray burst GRB 030329 has now
greatly advanced the issue of the GRB/SN association.
The GRB was detected by HETE-II at 2003 Mar 29.484 UT
\citep{van03} and its optical counterpart was promptly identified
\citep{pet03,tor03}. In the afterglow spectrum a bright Type Ic
SN~2003dh had been found \citep{sta03,cho03,zar03}. 
This detection is probably the conclusive confirmation that 
indeed some SNe and some GRBs are produced by the same 
astrophysical cauldron. Its unusual brightness provided us 
with a unique opportunity to obtain
spectroscopic and spectropolarimetric data of the SN about 40 days
after the GRB. Polarimetry is effective for proving the asymmetric
geometry of the explosion (e.g., \citealt{jef91,hoe91,kaw02}).

In this paper, we report on observations obtained with the Subaru
Telescope, and attempt to derive some of the properties of
SN~2003dh based on a comparison with those of other Type Ic
hypernovae.

\section{OBSERVATIONS}

The spectroscopic and spectropolarimetric observations of
SN~2003dh/GRB~030329 were obtained with FOCAS (Faint Object Camera
and Spectrograph; \citealt{kas02}) attached to the 8.2 m Subaru
Telescope on Mauna Kea. All the observations were carried out
through a polarimetric unit which consists of a rotating
superachromatic half-wave plate and a quartz Wollaston prism.
Spectra are obtained at four waveplate position angles 
($0^{\circ},\,45^{\circ},\,22\fdg 5
\mbox{ and } 67\fdg 5$) for spectropolarimetry or at one angle
($0^{\circ}$) for spectroscopy. A $0\farcs 8$ width slit and two 300
lines mm$^{-1}$ grisms were used: They produce 8 pixel $\sim$11
\AA\ resolution in both 5900-10240 \AA\ (red) and 4200-6000 \AA\
(blue). The spectropolarimetric data were obtained with the red
grism on 2003 May 8.3 and 9.3 UT, and their total exposure times
were 9120 s and 7200 s, respectively. The spectroscopic data were
obtained with the blue grism on May 9.4 with a 1200 s exposure
time. These epochs correspond to 34--35 rest-frame days after the
GRB assuming a redshift of $z=0.1685$ \citep{gre03}.

The data were reduced in the standard manner. Nightly observation
for unpolarized stars suggested that the
instrumental polarization was negligible ($p\lesssim 0.1$ \%), so
we did not correct for it. The PA of the polarization was 
corrected with dome flatfield spectra obtained
through a polarizing filter and using the observation of a
polarized star, Hiltner 960.

For flux calibration, we observed spectrophotometric standard star
Feige 34 \citep{mas90} with a wide ($2\farcs 0$) slit in a
moderate seeing condition ($1\farcs 0$).
In the target observation the $0\farcs 8$ width of the slit was 
comparable to or narrower than the target image sizes 
(FWHM$=1\farcs 0$-$1\farcs 6$ on May 8 and $0\farcs 7-0\farcs 8$ on May 9),
causing a non-negligible loss of light.
We estimated it from stellar image profiles in unfiltered
images taken for telescope guiding. It was approximately $39$
\% (May 8) or $24$ \% (May 9) and we corrected for it by
artificially scaling up the flux. 
The absolute value of the flux, however, is still lower
than \citet{zha03}'s result of $R=20.88$ on May 8.15 UT
by 0.45-0.50 mag.
The discrepancy could originate from an under-estimation of
the PSF of the stellar image or insufficient calibrations,
but the situation is unclear and we have not performed
any further correction.
Since the Galactic extinction toward the star is likely to be
relatively negligible ($A_{V}=0.065$; \citealt{sch98}),
we did not correct for it either.

In Figure \ref{fig1} we show the derived spectra of 
SN~2003dh/GRB~030329 in which several narrow emission lines are 
visible which can be attributed to \ion{H}{2} region within 
the host galaxy. There is no significant
variation in the supernova spectra over the two days.

\section{SPECTRAL COMPARISON WITH OTHER HYPERNOVAE}

\citet{sta03} compared the early spectrum of SN~2003dh with those
of SN~1998bw and found a close resemblance between them. They
pointed out that the expansion velocities in 1998bw and 2003dh on
Apr 8 are significantly higher than in other hypernovae, such as
SN~1997ef \citep{iwa00} and SN~2002ap \citep{maz02,kin02}.

Using our spectra, we performed a similar comparison.
We find that the spectra of SN~1998bw obtained between 19 and 29
days after $B$ maximum \citep{pat01}, 
i.e., between 33 and 43 rest frame days after
the explosion of the SN, are a good match for the May spectra of
SN~2003dh (Figure \ref{fig2}a). However, an even better match is
obtained using the spectra of SN~1997ef at a similar epoch (34--42
days; Figure \ref{fig2}b). In particular, the spectra of SN~2003dh
are different from those of SN~1998bw in the red, where in
SN~1998bw the \ion{O}{1} $\lambda 7774$ and \ion{Ca}{2} IR triplet
lines blend, while in SN~1997ef they do not. This is an indication
that SN~1997ef is less energetic than SN~1998bw, as the velocities
attained are smaller. Thus SN~2003dh on May 8--9 may also be less
energetic than SN~1998bw, although SN~2003dh had shown a
comparable expansion velocity in its early phase \citep{sta03}. Based
on the comparisons of earlier and later spectra, the total
kinetic energy of SN~2003dh may therefore lie between those of
SNe~1997ef and 1998bw. 

Such comparisons are quantitatively somewhat uncertain.
However, in this case they suggest that SN~2003dh exploded in the
range ($-8$,\,$+2$) days of GRB~030329, based on the similarity in
the spectral evolution to both SNe 1998bw and 1997ef.

Although, in the direct comparison with SNe~1997ef and 1998bw, 
SN 2003dh has a continuum excess in the blue region
($\lambda\lesssim 5000$ \AA ; see Figures \ref{fig2}a,b), 
it can be naturally explained by an underlying host galaxy 
and/or a residual afterglow continuum.
For example, using $\chi^{2}$-fitting to SN~1998bw,
we can plausibly extract the 
SN spectrum having little or no excess 
under the assumption that the host galaxy type 
is a starburst as shown in Figure \ref{fig3}.
(In this Letter, we use only the SN~1998bw spectrum 
for a fit because of much lower contamination in
the 1998bw host galaxy than in the 1997ef host galaxy.
Further analyses also using SN~1997ef spectra will appear
in future work.)
This seems almost consistent with the result in which an 
Sc-type galaxy template (assuming an LMC or SMC like dwarf 
galaxy of $V=22.7\pm 0.3$; \citealt{era03,fru03}; 
cf. Figure \ref{fig1}a) is subtracted from the observation 
(Figure \ref{fig4}).
In the decomposed spectrum, the \ion{Si}{2} $\lambda 6347,\,6371$ 
line in SN~2003dh appears to be comparable to the corresponding
feature in both SNe~1997ef and 1998bw.
Figure \ref{fig4} also suggests that the \ion{O}{1}
$\lambda 7774$ line in SN~2003dh is comparable to that
in SN~1997ef. Since O dominates the composition of the 
ejecta (see \citealt{nom01,nom03,nak01}), this may be taken to 
imply that the mass of the ejecta of SN~2003dh matches in 
SN~1997ef ($\sim 10$ M$_{\odot}$; \citealt{maz00}).

\section{POLARIMETRY}

The result of our polarimetry is shown in Figure \ref{fig1}c and
d. In the rest wavelength range 5000--7000 \AA, it is found
that the observed polarization is not significant ($P< 1$ \%),
although the alignment of PA around $40^{\circ}$ suggests that the
polarization is nonzero.
In the interstellar polarization catalogue \citep{hei00}
30 stars are found within 15$^{\circ}$ around SN~2003dh.
They mostly show only $P\lesssim 0.1$ \%.
The maximum polarization is found to be $P=0.2$ \% for HD 87737
($d=456$ pc) and the most distant star,
HD 91316 ($d=792$ pc), shows $P=0.17$ \%.
This is consistent with the Galactic extinction toward the SN
($A_{V}=0.065$) and the well-established relation of
$P_{\rm max}\simeq 3.1 A_{V}$ \citep{ser75}.
The interstellar polarization in our Galaxy is likely to be
negligible compared with the error in our data.

Various authors have reported significant polarization of the
GRB 030329/SN~2003dh object at the earliest stages.
The degree of polarization and its
position angle in $R$-band (center wavelength
corresponding to about 5560 \AA\ in rest frame ) were $0.90\pm 0.39$ \% and
$109^{\circ}\pm 12^{\circ}$ on Mar 30.8 UT \citep{efi03}, $1.97\pm 0.48$
\% and $83\fdg 2$ on Mar 31.1 \citep{mag03}, and $0.5\pm 0.1$
\% and $73^{\circ}\pm 5$ on Apr 2.1 \citep{cov03}. In this period,
the optical flux was dominated not by the supernova component but
by the GRB afterglow, and the variable polarization could be
attributed to the asymmetric fireball scenario (e.g.,
\citealt{ghi99}). However, the gradual rotation of PA with
wavelength found by \citet{cov03} suggests the existence of a
mildly polarized ($P\lesssim 0.5$ \%) component having
PA$\simeq$0--73$^{\circ}$, which is consistent with our value $\sim
40^{\circ}$. This is consistent with a constant
polarization component due to interstellar matter in the
host galaxy (dichroic absorption and/or scattering).
The similarity of the polarization PA of
our data at 5000--7000 \AA\, and the PA of elongation of the
host galaxy ($230^{\circ}=50^{\circ}$; \citealt{fru03}), also 
suggest that a considerable fraction of the polarization is due to
interstellar polarization in the host galaxy.

The low intrinsic polarization of SN~2003dh can be explained in
terms of a mild asphericity of the explosion along the line of
sight (e.g., \citealt{hoe91}), but this does not imply that the
explosion was highly spherically symmetric.  If the explosion is
spheroidal (e.g., prolate) and we view it nearly along the axis of
the symmetry, only a small departure from asymmetry about
the line of sight is expected, and thus only small polarization
is expected. It has been known that core-collapse SNe
usually show large polarization, and their explosions are likely
to be asymmetric \citep{wan96,wan01,leo01}. The presence of the
GRB and the hypernova spectrum suggests that the viewing angle
nearly aligns with the axis of the prolate explosion as 
suggested for SN 1998bw/GRB 980425 
(e.g., \citealt{hoe99,pat01,mae02}).

We also should note the possible polarization at 7000--8000
\AA , which has a PA orthogonal to that of the polarization at
wavelengths 5000--7000 \AA . SN 2002ap, which is at present the
only hypernova that has been well observed with
spectropolarimetry, also had distinct polarization features
($\Delta P\sim$ 1--2 \%) at \ion{O}{1} or \ion{Ca}{2} triplet
absorption troughs \citep{kaw02,leo02,wan03}. However, Figure
\ref{fig1}c shows that the significance of the polarization at
7000--8000 \AA\ is low ($P/\sigma_{P} < 2$). The contamination of
residual sky emission lines could degrade the quality in the
wavelength region. We conclude that the significance of the
polarization in SN~2003dh on May 8 and 9 is not high compared
with other core-collapse SNe.

\section{CONCLUSIONS}

Our observations confirm that the supernova component SN~2003dh of
GRB~030329 evolves in the same manner as other Type Ic hypernovae,
e.g., SNe~1997ef and SN~1998bw.
Polarization was marginally detected
for the SN at optical wavelengths, but it seems to be
mainly (perhaps totally) of interstellar origin in the host
galaxy. The low intrinsic polarization would be consistent with
the scenario of the highly-collimated jet model of the GRB
afterglow if the viewing angle is well aligned with the axis
of asymmetric explosion. 
However, this suggestion is tentative. Future bright 
SNe/GRBs should also be monitored spectropolarimetrically.

\acknowledgements
We are grateful to the staff of the Subaru Telescope for
their kind support.
This work has been supported in part by the grant-in-Aid for
Scientific Research (14047206, 14540223, 1520401) of the Ministry of 
Education, Science, Culture, Sports, and Technology in Japan.

\figcaption[fig1.eps]{
Observations of SN~2003dh/GRB~030329:
(a) Flux spectra in which the telluric absorption
has been corrected for, (b) spectra of sky emission
lines and telluric absorptions for a reference of
less-reliable wavelength regions,
(c) degree of observed polarization, and (d)
its position angle in the equatorial coordinate.
The flux has been binned to $10.8$ \AA\ widths.
In panel (a), a template spectrum of an Sc-type
galaxy (GISSEL98; \citealt{bru93}) of $V=22.7$ \citep{zha03} 
is shown as well as 
a template GRB continuum having power-low index
$\beta=-0.9$ \citep{sta03} for comparison.
The polarimetric data are properly binned and the each
band width is indicated by horizontal error bar.
The vertical error bar denotes its mean error ($1\sigma$).
\label{fig1}}

\figcaption[fig2a.eps,fig2b.eps]{
Comparison of SN~2003dh spectrum with other hypernovae: (a)
SN~1998bw \citep{pat01} and (b) SN~1997ef \citep{maz00}.
The SN~2003dh spectrum is an average of
both May 8 and 9 spectra.
These show a global resemblance between the spectrum
of SN~2003dh and those of the other hypernovae.
However, the match seems better with SN~1997ef around
the \ion{O}{1} $\lambda$ 7774 line.
\label{fig2}}

\figcaption[fig3.eps]{
SN~2003dh spectrum after subtraction of a starburst galaxy spectrum.
They are derived using $\chi^{2}$-fitting to SN~1998bw spectrum 
36 days after explosion.
The thick dashed line shows the best matching SN~1998bw spectrum,
the dash-dotted line is the subtracted host spectrum, and
the continuous line is the decomposed SN~2003dh spectrum. 
\label{fig3}}

\figcaption[fig4.eps]{
Comparison between the SN~2003dh spectrum from which a 
template spectrum of an Sc-type galaxy (cf. Figure \ref{fig1})
has been subtracted and those of other hypernovae.
\label{fig4}}


\plotone{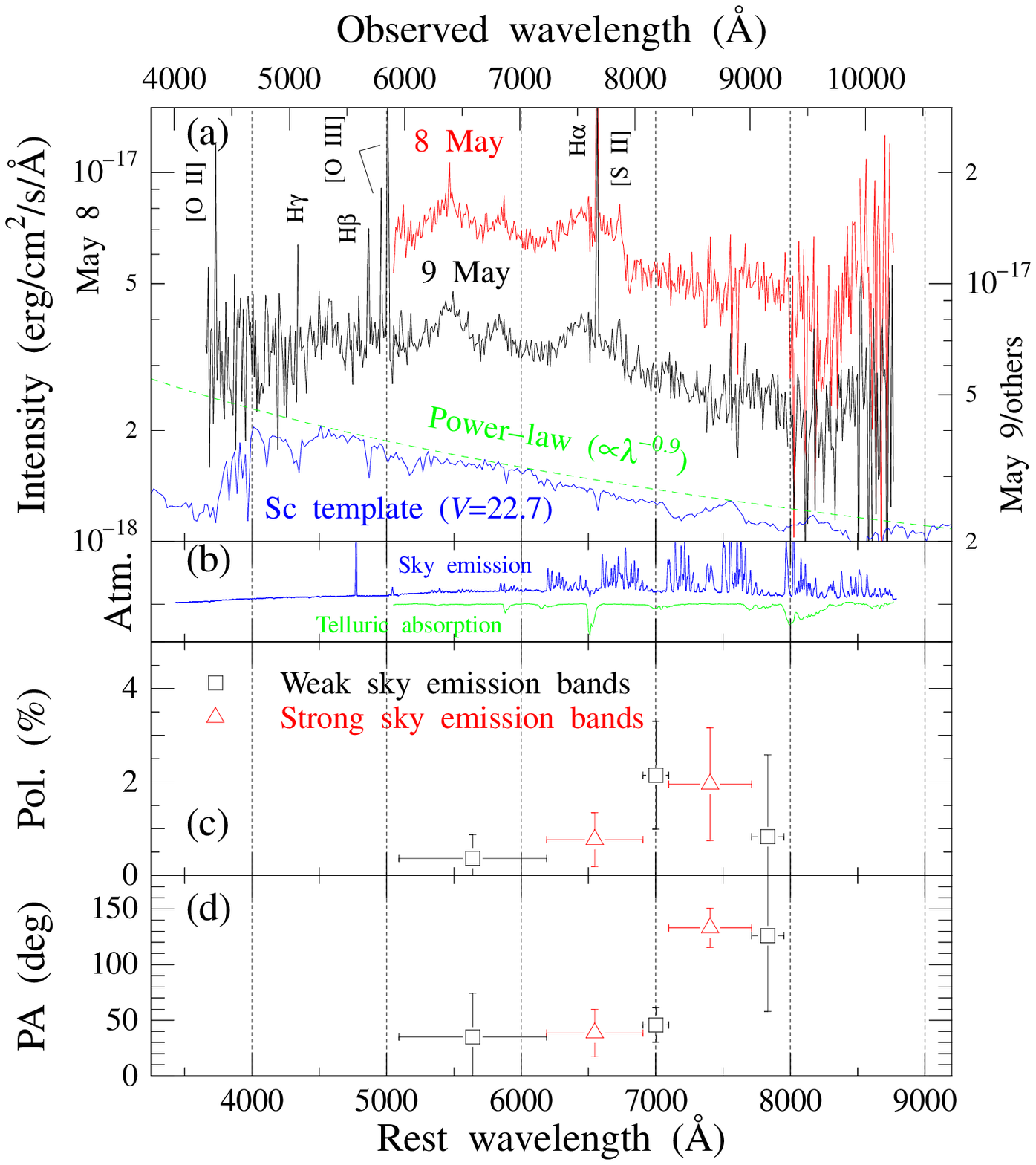}

\plotone{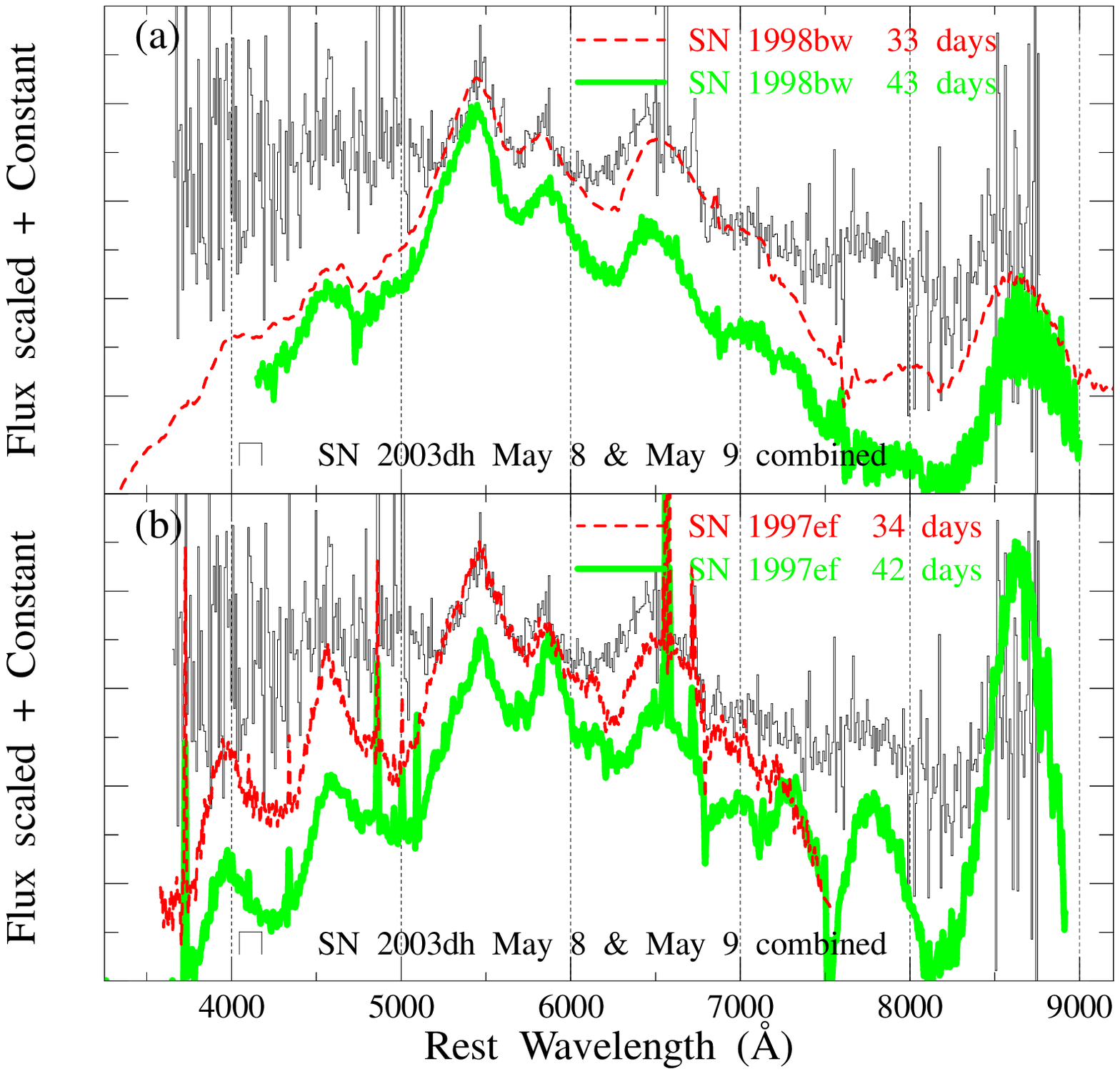}

\plotone{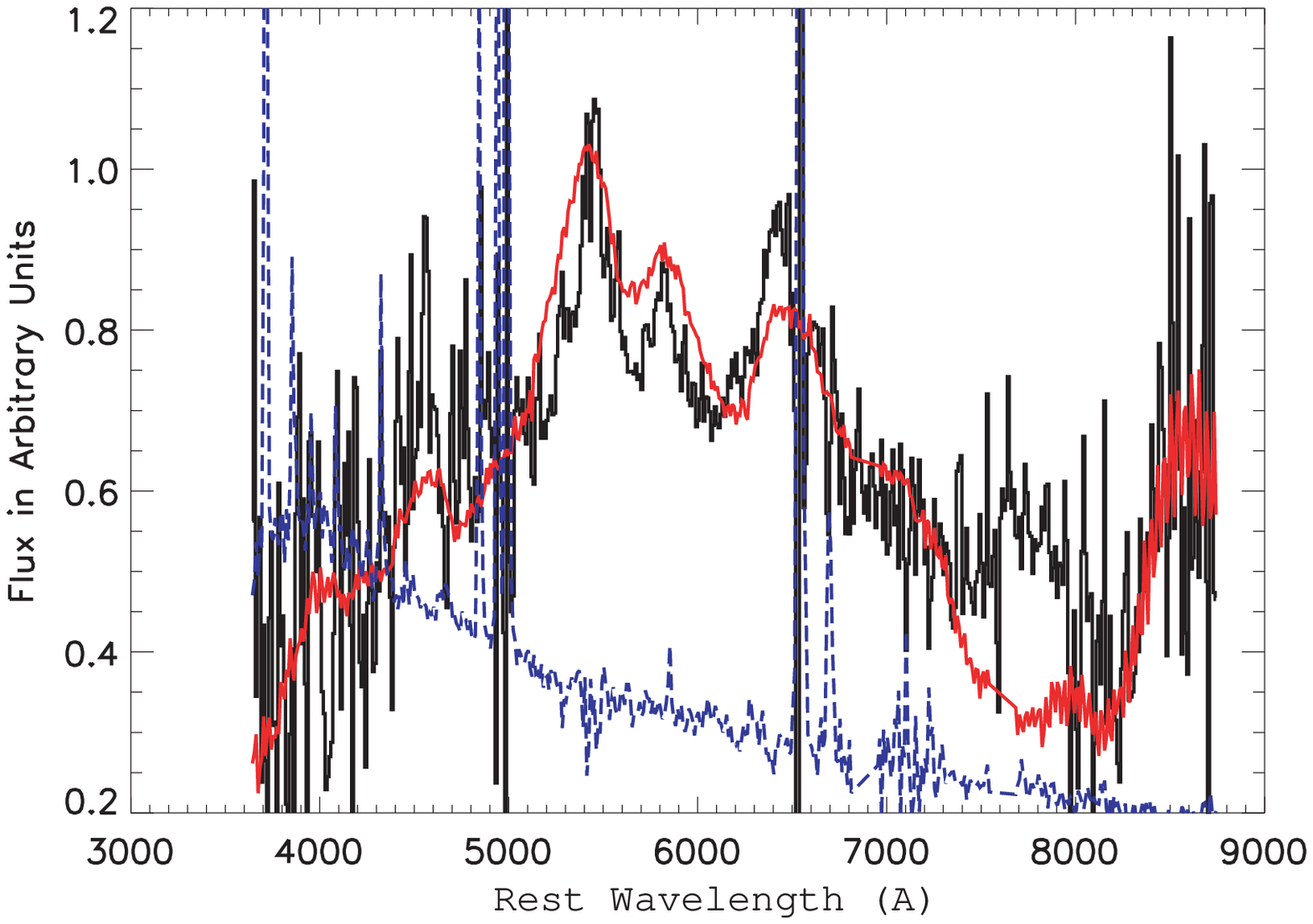}

\plotone{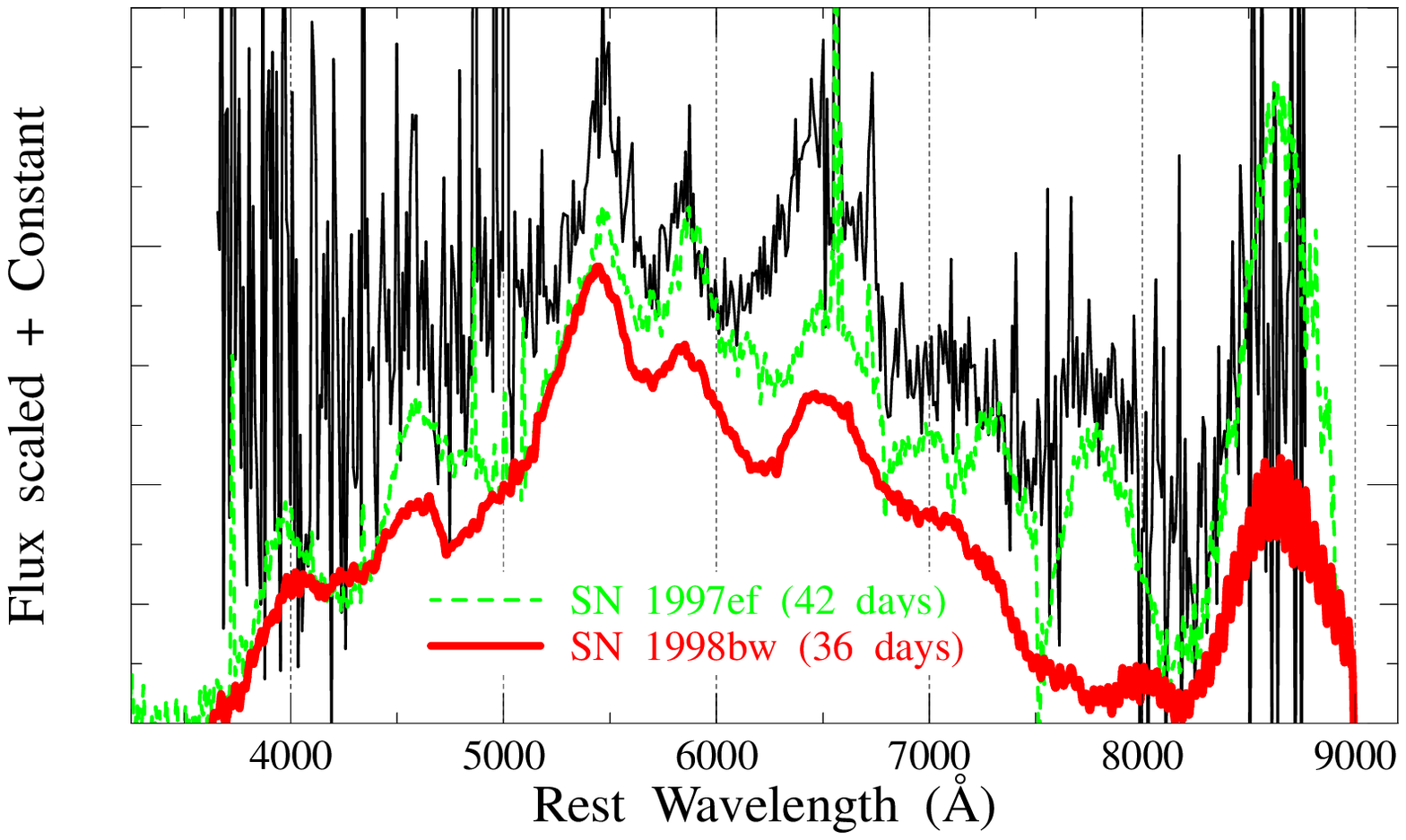}

\end{document}